\documentstyle[12pt,epsf]{article}
\newcommand{\nll}{\nonumber \\}

\newcommand{\bq}{\begin{equation}}
\newcommand{\eq}{\end{equation}}
\newcommand{\ba}{\begin{eqnarray}}
\newcommand{\ea}{\end{eqnarray}}

\newcommand{\nobody}{\rule{0ex}{1ex}}

\newcommand{\vc}{\char'24\hspace{-1ex}c}
\begin{document}
\voffset -2cm
\hfill Interner Bericht

\hfill DESY-Zeuthen 96--06

\hfill LMU--01/96

\hfill March 1996

\hfill hep-ph/9604321

\begin{center}
\vspace{1.5cm}\hfill\\ 
{\LARGE  $Z'$ Constraints from $e^+e^-\rightarrow f\bar f$ at NLC
\footnote{To appear in {\it Proceedings of Physics with e$^+$e$^-$ Linear
Colliders Workshop}, Annecy -- Gran Sasso -- Hamburg 1995, ed. P. Zerwas}
}
\vspace{1cm}\\
A. Leike$^a$, S. Riemann$^b$\\
$\nobody^a$ {\it
Ludwigs--Maximilians-Universit\"at, Sektion Physik, Theresienstr. 37,\\
D-80333 M\"unchen, Germany}\\
E-mail: leike@graviton.hep.physik.uni-muenchen.de\\
$\nobody^b$ {\it
Deutsches Elektronen-Synchrotron DESY\\
Institut f\"ur Hochenergiephysik IfH, Zeuthen,\\
 Platanenallee 6, D-15738 Zeuthen,
Germany}\\
E-mail: riemanns@ifh.de\\
\end{center}
\begin{abstract}
\noindent
Constraints on extra neutral gauge bosons are obtained from 
$e^+e^-\rightarrow f\bar f$ at NLC energies.
Model independent limits on the $Z'f\bar f$ couplings
and lower limits on the $Z'$ mass
are discussed.
Typical GUTs with $M_{Z'}$ up to (3 -- 6) $\sqrt{s}$
can be excluded.
A distinction between different GUT scenarios
is possible if  $M_{Z'}\le 3\sqrt{s}$.
Radiative corrections give only small changes to the 
$Z'$ exclusion limits provided that appropriate cuts are applied.
 \vspace{1cm}
\end{abstract}
%
\section{Introduction}
The search for extra neutral gauge bosons is an important task of the 
physics programme of all present and future colliders.
Up to now, no $Z'$ signals are found. 
These experimental results are usually reported as lower
limits on excluded $Z'$ masses or as upper limits on the $ZZ'$
mixing angle for selected $Z'$ models. 
With future colliders one can find a $Z'$ or
put much stronger constraints, see
\cite{zpsari,zpmi,delaguila}.

In this paper and in a more detailed analysis \cite{alsr},
we examine the $Z'$ constraints which can be obtained
from $e^+e^-\rightarrow f\bar f$ at c.m. energies
$\sqrt{s}=500\,GeV$ with $L_{int}=20fb^{-1}$ (NLC500) or
$\sqrt{s}=2\,TeV$ with $L_{int}=320fb^{-1}$ (NLC2000)
and $P=80$\% polarization of the $e^-$ beam in both scenarios.
In comparison to \cite{zpsari,zpmi,delaguila}, we take into account all 
available QED, electroweak and QCD corrections and apply kinematical
cuts in order  to approach a more realistic description of future detectors.
Including more observables into our analysis,
we go beyond \cite{zpsari,zpmi}.
In contrast to \cite{delaguila}, we additionally include the expected
systematic errors.

We set the $ZZ'$ mixing angle zero in accordance with present
experimental constraints  \cite{zmix,zmixl3,zefit}.
CDF data indicate that NLC500 will operate below a $Z'$ peak
\cite{cdfzp94}. 
Similarly, LHC will be able to exclude 
a $Z'$ which could be produced at NLC2000 on resonance. 
We assume that NLC2000 will operate below the $Z'$ peak, too.
Throughout we presume universality of generations.
Theories including extra neutral gauge bosons usually predict  new
fermions \cite{maa,e6}. Their effects are neglected here.

We consider $Z'$ models which are described by the following effective 
Lagrangian at low energies, 
\begin{equation}
{\cal L} = 
e A_\beta J^\beta_\gamma + g_1 Z_\beta J^\beta_Z + g_2 Z'_\beta J^\beta_{Z'}.
\end{equation}
The term proportional to $g_2$ contains the new
interactions of the $Z'$ with Standard Model fermions.

Although we mainly focus on model independent $Z'$ limits, we will
also refer to some special models predicted in an $E_6$ GUT
\cite{e6,e6lr} and in a Left-Right Model \cite{e6lr,lr},
\begin{equation}
\label{zeparm}
J^ \mu_{Z'} = J^\mu_\chi \cos\beta + J^\mu_\psi \sin\beta,\ \ \ 
J^\mu_{Z'} = \alpha_{LR} J^\mu_{3R} - \frac{1}{2\alpha_{LR}} J^\mu_{B-L}.
\end{equation}
Some completely specified cases are $Z'=\chi,\ \psi$ and 
$\eta$ $\left(\beta=-\arctan\sqrt{5/3}\right)$ in the $E_6$ GUT.
Special cases in the Left-Right Model are obtained for 
$\alpha_{LR}$ equal to $\sqrt{2/3}$ and $\sqrt{\cot^2{\theta_W}-1}$. 
The first value of $\alpha_{LR}$ gives again the $\chi$ model, while the 
second number gives the Left-Right Symmetric Model (LR). 
The Sequential Standard Model (SSM) is also considered.
It contains a heavy $Z'$  whith exactly the same couplings to fermions
as  the Standard $Z$ boson.

The presence of an extra neutral gauge boson leads to an additional 
amplitude of fermion pair production at the Born level,
\begin{eqnarray}
\label{link}
{\cal M}(Z')  = \frac{g_2^2}{s-m_{Z'}^2}                            
\bar{u}_e\gamma_\beta (\gamma_5 a'_e + v'_e ) u_e \, 
\bar{u}_f\gamma^\beta (\gamma_5 a'_f + v'_f ) u_f
\hspace{5cm} 
\\ \nonumber 
= 
 -\frac{4 \pi}{s} \left[
\bar{u}_e\gamma_\beta (\gamma_5 a_e^N + v_e^N ) u_e \, 
\bar{u}_f\gamma^\beta (\gamma_5 a_f^N + v_f^N ) u_f \right]
\hspace{5cm} 
\end{eqnarray}
\begin{eqnarray}
\label{link2}
\mbox{\ with\ } 
a_f^N = a'_f \sqrt{\frac{g_2^2}{4 \pi} \frac{s}{m_{Z'}^2-s}}\ ,\ 
v_f^N = v'_f \sqrt{\frac{g_2^2}{4 \pi} \frac{s}{m_{Z'}^2-s}},\
\mbox{and}\ \ m_{Z'}^2 = M_{Z'}^2-i\Gamma_{Z'} M_{Z'}.
\end{eqnarray}
Equations (\ref{link}) and (\ref{link2}) show that 
far below the resonance
the effect of a $Z'$ is described by the two parameters  $a_f^N$ and $v_f^N$ 
and not by $a'_f,\ v'_f$ and $m_{Z'}$ separately. 
The fermionic couplings of some $Z'$
define a point in the $(a_f^N,\ v_f^N)$ planes ($f=\nu,l,u,d$). 
Various observables can detect a $Z'$ in different regions of the  
$(a_f^N,v_f^N)$ planes. 

The $e^+e^-$ colliders provide several observables
depending on couplings to only leptons as
\bq
\label{obs}
\sigma_t^l,\ A_{FB}^l,\ A_{LR}^l,\ A_{pol}^\tau
,\ A_{pol,FB}^\tau\mbox{\ and\ } A_{LR,FB}^l.
\eq
Therefore, the $Z'$ couplings to leptons $(a_l^N,v_l^N)$ can be
constrained independently of the quar\-ko\-nic $Z'$ couplings.
The index $l$ stands for electrons and muons the final state.
Only the $s$ channel is considered for electrons in the final state.
Neglecting fermion masses, the last four
leptonic observables in (\ref{obs}) are related at the Born level,
\bq
A_{LR}^l = A_{pol}^\tau = \frac{4}{3} A_{pol,FB}^\tau =\frac{4}{3}A_{LR,FB}^l.
\eq
Therefore, they are equivalent for a $Z'$ search. Without loss of
generality, we will consider only $A_{LR}$ as a representative. 
It is expected to have the smallest error compared to the other three
observables. 

The hadronic observables can be divided into three groups:
Observables containing in addition to the obligatory $Z'l\bar l$ couplings 
the $Z'b\bar b$ couplings only, as
\bq
R_b=\sigma_t^b/\sigma_t^\mu,\ A_{FB}^b,\ A_{LR}^b,
\eq
the $Z'c\bar c$ couplings only, and all couplings of the $Z'$ to
quarks. 
The analysis of the first two groups is very similar. 
However, we will not consider the second group because
$c$-quark flavour identification leads to systematic errors which are
considerably larger than those from $b$-quark identification.
The third group has the smallest errors because it doesn't require flavour
identification.
We consider
\bq 
R^{had}=\sigma_t^{had}/\sigma_t^\mu=\sigma_t^{u+d+s+c+b}/\sigma_t^\mu
\mbox{\ and\ }
A_{LR}^{had}=A_{LR}^{u+d+s+c+b}.
\eq

The statistical errors of all observables for $N$ detected events are 
\begin{equation}
\label{stat}
\frac{\Delta\sigma_t}{\sigma_t}=\frac{1}{\sqrt{N}},\ \ \  
\Delta A_{FB} = \sqrt{\frac{1-A^2}{N}},\ \ \ 
\Delta A_{LR} = \sqrt{ \frac{1 -(P A_{LR})^2}{N P^2} }\ .
\end{equation}
We assume a systematic error of the luminosity measurement of 0.5\% .
Further, we include a systematic error of 0.5\% for the measurement of each
observable.
We assume 1\% systematic error due to $b$-quark identification
and take into account the efficiency of quark tagging.
Statistical and systematic errors are added in quadrature.
The resulting combined errors are equal for NLC500 and NLC2000:
\ba
\label{comberr}
\Delta\sigma_t^l/\sigma_t^l=1\%,\ 
\Delta A_{FB}^l=1\%,\ 
\Delta A_{LR}^l=1.2\%,\nll
\Delta R_b=2.2\%,\
\Delta A_{FB}^b=2.0\%,\
\Delta A_{LR}^b=1.5\%,\
\Delta R^{had}=0.9\%,
\Delta A_{LR}^{had}=0.7\%.
\ea

To obtain confidence levels for different sets of parameters, 
we calculate the prediction for all
observables in the Standard Model
$O_i($SM$)$ and in a theory including a $Z'$
$O_i($SM$,v_l^N,a_l^N)$ and consider the deviation of
\bq
\chi^2 = \sum_{O_i}
\left[\frac{O_i(\mbox{SM})-O_i(\mbox{SM},v_l^N,a_l^N)}{\Delta O_i}
\right]^2 
\eq
from the minimum.

With our assumptions, the $Z'$ can be detected only through small deviations
of observables from their Standard Model  predictions.
It is known that radiative corrections have to be included to meet the
expected experimental precision.
For all energies considered here, the QED corrections are numerically most
important. 
The energy spectrum of the radiated photons has a huge peak for 
energies $E_\gamma/E_{beam}\approx 1-M_Z^2/s$. 
This is due to the radiative return to the $Z$ resonance. 
Such events do not contain information about new heavy particles.
They may  be eliminated by a cut on hard photons.
This can be realized by a cut on the photon energy, 
$E_\gamma/E_{beam} = \Delta < 1-M_Z^2/s$, or by a cut
on the acollinearity angle of the two outgoing fermions. 
Satisfying these conditions, the analysis is much less sensitive to further
cuts.

\section{Model independent $Z'$ limits}
Our analysis is performed by the code {\tt ZEFIT} 
\cite{zefit} which works together with  {\tt ZFITTER} \cite{zfitter}.
All Standard Model corrections and all possibilities
to apply kinematical cuts available in {\tt ZFITTER} are made operative.
The code {\tt ZEFIT} contains the additional $Z'$ contributions.
It was already applied to LEP\,1 data to set bounds to the $ZZ'$ mixing angle 
\cite{zmixl3} and is now adapted to a model independent $Z'$ analysis.
QED corrections to the new $Z'$ interferences are applied to the same order
as to the SM cross section. 
In our fits, we used the full one-loop electroweak corrections, the 
QCD corrections and soft photon
exponentiation for photons in the initial and final states. 
The initial state radiation was taken at two-loops.
We forbid hard photons taking
$\Delta=0.9$ for NLC500 and $\Delta=0.98$ for NLC2000. 
As a simple simulation of the detector acceptance, 
we demand that the angle between the outgoing
leptons and the beam axis is larger than $20^\circ$. We apply no angular
restrictions to outgoing quarks. 
Possible correlations between the errors of different observables 
are neglected.

\begin{figure}
\begin{minipage}[t]{7.8cm}{
\begin{center}
\hspace{-1.7cm}
\mbox{
\epsfysize=7.8cm
\mbox{\epsfxsize=7.8cm\epsffile{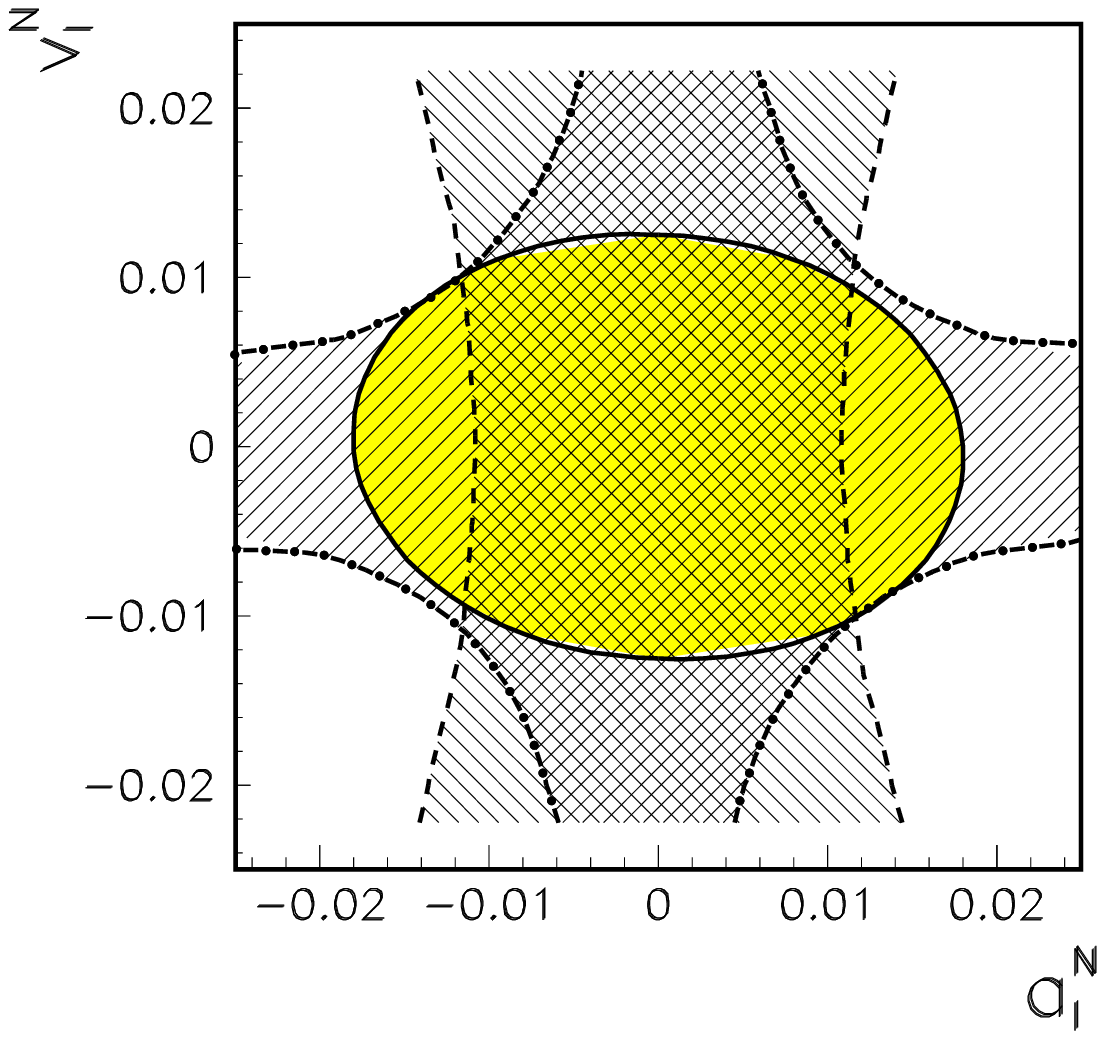}}
}
\end{center}
\noindent
{\small\bf Fig.~1: }{\small\it
Areas in the $(a_l^N,\ v_l^N)$ plane indistinguishable from the SM
at NLC500 (95\% C.L.).
$\sigma_t^l$ cannot distinguish models inside the shaded ellipse.
$A_{FB}^l$ ($A_{LR}^l$) are blind to models inside the hatched areas with 
falling (rising) lines. The region from all observables
combined is also shown (thick line).
}
}\end{minipage}
\hspace{0.5cm}
\begin{minipage}[t]{7.8cm}{
\begin{center}
\hspace{-1.7cm}
\mbox{
\epsfysize=7.8cm
\mbox{\epsfxsize=7.8cm\epsffile{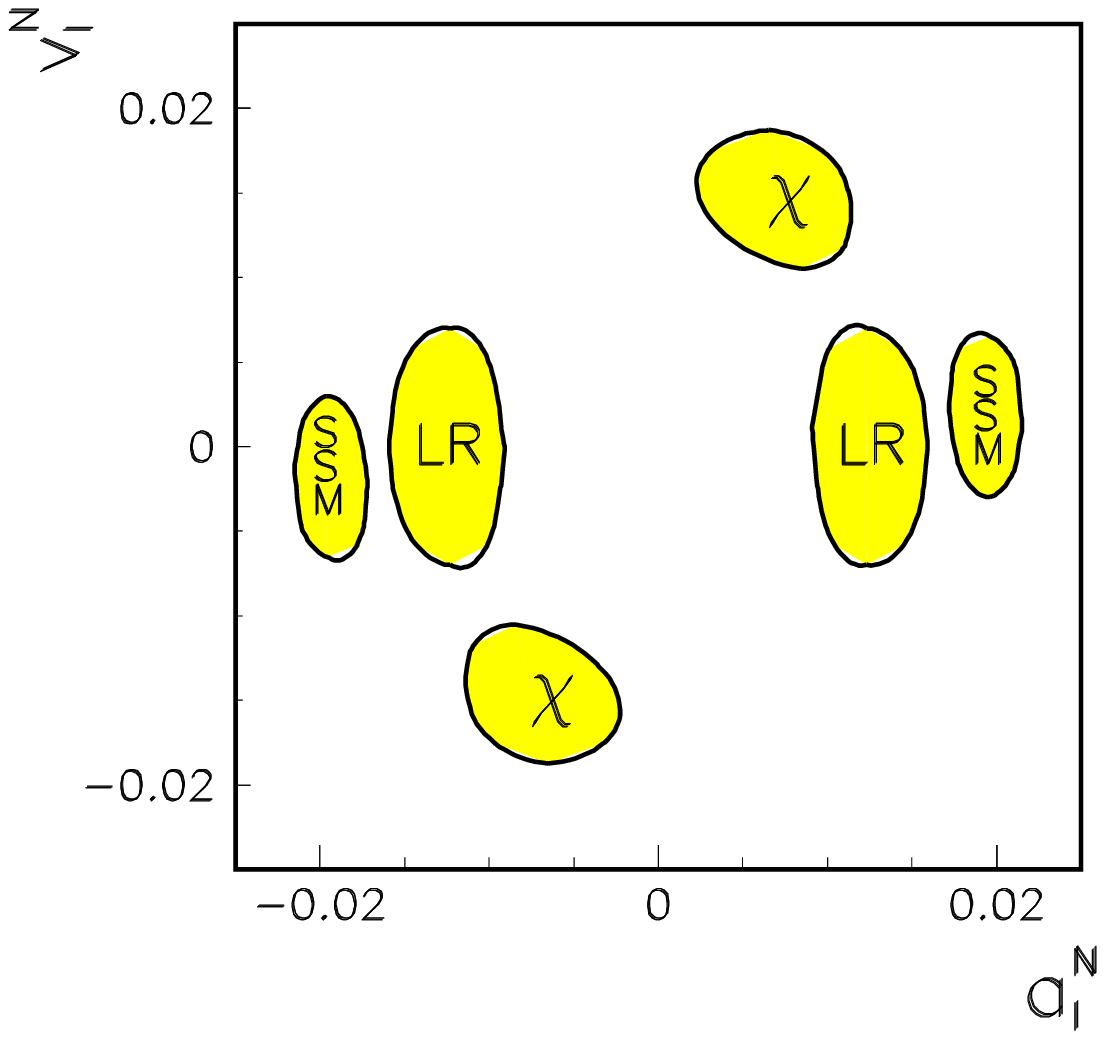}}
}
\end{center}
\noindent {\small\bf Fig.~2: }{\small\it 
Resolution power of NLC500 
(95\% C.L.)
by all leptonic observables combined for different models and
$M_{Z'}=3\sqrt{s}=1.5\,TeV$.
}
}\end{minipage}
\end{figure}
%

Figure~1 shows our model independent discovery limits on the couplings of
the $Z'$ to leptons from different observables.
The corresponding figures for NLC500 and NLC2000 are 
almost identical because 
we expect the same number of events for both collider scenarios
and $a_l^N$ and $v_l^N$ are normalized.
We see that $\sigma_t^l$ constrains both the vector and the axial
vector couplings, 
while $A_{FB}^l$ constrains mainly the axial vector couplings.
$A_{LR}^l$ gives only a minor improvement to the discovery limits.
The region obtained by all leptonic observables in combination  is also shown.
The changes of figure~1 for experimental 
errors different from our assumptions are described by the Born
formulae given in ref. \cite{zpmi}.
 
Assuming the existence of a Z$'$,   one can
distinguish between different models at NLC500.
This is shown in figure~2 for $M_{Z'}=3\sqrt{s}$.
The corresponding figure for NLC2000 
does not differ from figure~2.
In contrast to figure~1, $A_{LR}^l$
is a really important input because it
is sensitive to the sign of the $Z'$ couplings.
Note that a simultaneous change of the sign of both leptonic $Z'$ couplings 
can never be detected by the reaction $e^+e^-\rightarrow f\bar f$.

The constraints on $a_l^N$ and $v_l^N$ by the three leptonic observables
could in principle lead to contradicting results.
This could happen, if an area allowed by two observables, e.g. $\sigma_t^l$
and $A_{FB}^l$ is excluded by a third observable (e.g. $A_{LR}^l$). Such a case
would be an indication for new physics beyond a $Z'$. 

Non-zero $Z'$ couplings  to leptons are necessary for a $Z'$ signal in
the hadronic observables. 
To be definite, we assume that the $Z'$ is described by one of the 
models $\chi$, LR or SSM considered in figure~2.
As in the leptonic case, different $b$-quark observables are blind in 
different directions. 
Polarized beams give a large improvement to the measurement of $Z'b\bar b$ 
couplings.
If combined, all $b$-quark observables define a closed region in the
$(a_b^N,v_b^N)$ plane. 
Figure~3 shows that for all three models
one can detect a non-zero $Z'$ signal also 
in the $Z'b\bar b$ couplings.
However, one cannot discriminate between $\chi$ and LR as 
it was possible in the leptonic sector.
We refer to \cite{alsr} for more details.

\begin{figure}
\begin{center}
\begin{minipage}[t]{7.8cm} {
\begin{center}
\hspace{-1.7cm} \mbox{ \epsfysize=7.0cm \epsffile[0 0 500 500]{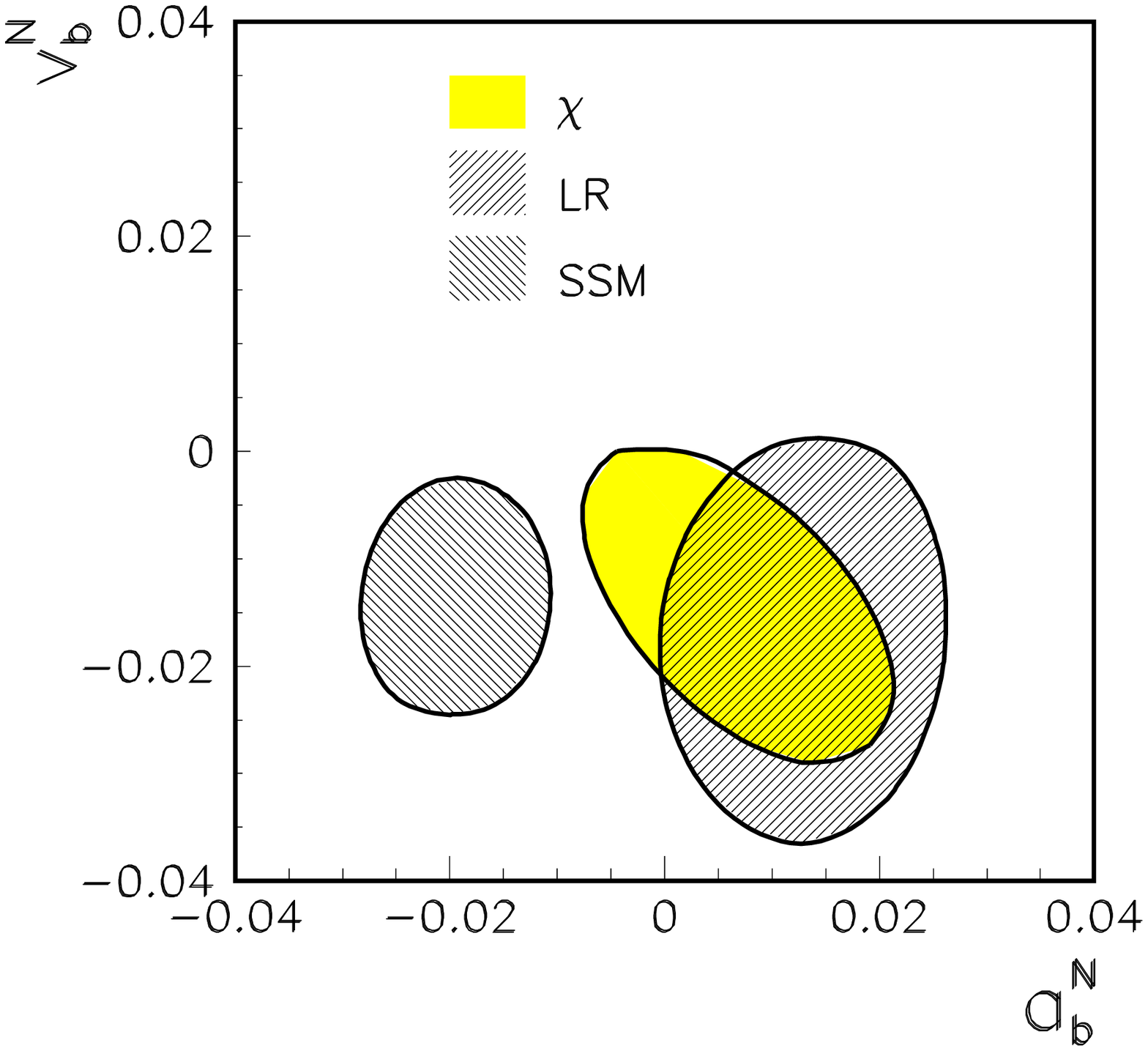}
}
\end{center}
}\end{minipage}
\end{center}
\noindent {\small\bf Fig.~3: }{\small\it 
Resolution power of NLC500 in the $(a_b^N,v_b^N)$ plane (95\% C.L.)
by all $b$-quark observables combined.
Different $Z'$ models with $M_{Z'}=1.5\,TeV$ are assumed. 
}
\end{figure}
%

\section{Model dependent $Z'$ limits}
We now discuss the results of several one parameter fits with and
without systematic errors.

The lower limits (95\% C.L.) on the $Z'$ mass, $M_{Z'}^{lim}$,
for different $Z'$ models are given in table~1.
The observables $R^{had}$ and $ A_{LR}^{had}$ give an important input
to these fits.
Table~1 shows three rows for every collider. The first row contains the
limits from leptonic observables only.
The second row includes leptonic and hadronic observables. 
The third row includes all observables as the second row but without 
systematic errors. 
Comparing the first two rows of a certain collider, we see that the
hadronic observables improve the mass limits between 5\% and 10\%.
Therefore,  both the leptonic and hadronic observables
are  important for measurements of $M_{Z'}^{lim}$.
The difference between the numbers of the 
second and third row is about 10\%. 
Hence, the mass limits obtained for a particular $Z'$ model are rather
insensitive to the assumptions about systematic errors.

The errors of hadronic observables are dominated 
by systematic uncertainties. Thus,
the relatively large vector and axialvector couplings
of the Z$'$ in the SSM
cause a larger sensitivity of  $M_{Z'}^{lim}$
to systematic errors.


As a result, $e^+e^-$ colliders can exclude a
$Z'$ with a mass lighter than $M_{Z'}^{lim}\sim 3$ to $6 \sqrt{s}$ 
for popular GUT's and  $M_{Z'}^{lim}\sim 8\sqrt{s}$ for the SSM.

\begin{table}
\begin{tabular}{lrrrrrr}\hline\\
  & & $\chi$ & $\psi$ & $\eta$ & LR & SSM\\ \hline\\[0.5ex] 
NLC500    &leptonic observables only & 2670 & 1900 & 1570 & 2700 & 3720 \\ 
          &all observables           & 2850 & 2000 & 1730 & 3470 & 4450 \\ 
&all observables no systematic errors& 3230 & 2160 & 2000 & 4040 & 5500 \\[2ex]
NLC2000   &leptonic observables only & 9500 & 6950 & 5780 & 9820 &13560 \\ 
          &all observables           & 9590 & 6980 & 6020 &10300 &14000 \\ 
&all observables no systematic errors&11200 & 7850 & 6840 &12700 &17800\\
\hline
\end{tabular}\medskip

{\small\it Tab.~1: Maximal $Z'$ masses $M_{Z'}^{lim}$ (in GeV) which
  can be excluded (95\% C.L.) by leptonic observables only and by all observables
  combined for NLC500 and NLC2000.  } 
\end{table} 

To summarize, we investigated the limits on extra neutral gauge bosons
which can be achieved at future linear $e^+e^-$ colliders. 
We included radiative corrections, kinematical cuts and systematic
errors in our analysis. The influence of radiative corrections on the
$Z'$ mass limits is small after applying appropriate cuts. 
Systematic errors have a moderate influence on the $Z'$ mass limits
but they are important to distinct models .

%
%


\begin{thebibliography}{99}
\bibitem{zpsari} A. Djouadi, A. Leike, T. Riemann, D. Schaile, C. Verzegnassi,
                    Proc. of the ``Workshop on Physics and Experiments with 
                    Linear Colliders'', Sept. 1991, Saariselk\"a, Finland, ed.
                    R. Orava, Vol. II, p. 515;\\
                 A. Djouadi, A. Leike, T. Riemann, D. Schaile, C. Verzegnassi,
                    Z. Phys. {\bf C56} (1992) 289.
\bibitem{zpmi} A. Leike, Z. Phys. {\bf C62} (1994) 265.
\bibitem{delaguila}  F.del Aguila, M. Cveti\vc, P. Langacker, 
                     Phys. Rev. {\bf D48} (1993) R969;\\
  F.del Aguila, M. Cveti\vc, Phys. Rev. {\bf D50} (1994) 3158;\\
  F.del Aguila, M. Cveti\vc, P. Langacker, Phys. Rev. {\bf D52} (1995) 37
\bibitem{alsr}   A. Leike, S. Riemann, in preparation.
\bibitem{zmix} P. Langacker, M. Luo, Phys. Rev. {\bf D45} (1992) 278;\\
                G. Altarelli, et al., Phys. Lett. {\bf B318} (1993) 139.
\bibitem{zmixl3}  L3 collaboration, Phys. Lett. {\bf B306} (1993) 187.
\bibitem{zefit} A. Leike, S. Riemann, T. Riemann, Phys. Lett. {\bf B291} 
                (1992) 187.
\bibitem{cdfzp94}
 L. Nodulman,  CDF Collab., New Particle Searches at CDF, to appear in
proceedings of the EPS Conference, Brussels, 1995.
\bibitem{maa} J. Maalampi, M. Roos, Phys. Rep. {\bf 186} (1990) 53.
\bibitem{e6} For a review see e.g.
             J.L. Hewett, T.G. Rizzo, Phys. Rep. {\bf 183} (1989) 193.
\bibitem{e6lr} L.S. Durkin, P. Langacker, Phys. Lett. {\bf B166} (1986) 436.
\bibitem{lr} For a review see e.g.
             R.N. Mohapatra, {\it Unifications and Supersymmetries}, Springer,
                New York 1989.
\bibitem{zfitter}
D. Bardin et al., FORTRAN package ZFITTER Version 4.8;\\
D. Bardin et al., Preprint CERN--TH/6443/92, hep-ph/9412201;\\
D. Bardin et al., Z. Phys. {\bf C44} (1989) 493;\\
D. Bardin et al., Nucl. Phys. {\bf B351} (1991) 1;\\
D. Bardin et al., Phys. Lett. {\bf B255} (1991) 290.
\bibitem{slr} A. Leike, T. Riemann, M. Sachwitz, Phys. Lett. {\bf B241} 
                (1990) 267;
                A. Leike, T. Riemann, Z. Phys. {\bf C51} (1991) 321.
\end{thebibliography}
\end{document}